\documentclass[longauth]{aa}
\usepackage{natbib}
\bibpunct{(}{)}{;}{a}{}{,} % to follow the A&A style

\usepackage{xcolor}
\usepackage{ulem}
\usepackage{amsmath}
\usepackage{amsfonts}
\usepackage{amssymb}
\usepackage[switch]{lineno}

\usepackage[varg]{txfonts}

\newcommand\LST{LST\nobreakdash-1}

\usepackage{graphicx}
 
\begin{document} 

%%%%%%%%%%%%%%%%%%%%%%%%%%%%%%%%%%%%%%%%%%%%%%%%%%%%%%%%%%%%%%%%%%%%%%%%%%%%%%%%%
%
% TITLE
%
\title{A detailed study of the very-high-energy Crab pulsar emission with the LST-1}

%%%%%%%%%%%%%%%%%%%%%%%%%%%%%%%%%%%%%%%%%%%%%%%%%%%%%%%%%%%%%%%%%%%%%%%%%%%%%%%%%
%
% AUTHORS
%
\author{
K.~Abe\inst{1} \and
S.~Abe\inst{2} \and
A.~Abhishek\inst{3} \and
F.~Acero\inst{4,5} \and
A.~Aguasca-Cabot\inst{6} \and
I.~Agudo\inst{7} \and
N.~Alvarez~Crespo\inst{8} \and
L.~A.~Antonelli\inst{9} \and
C.~Aramo\inst{10} \and
A.~Arbet-Engels\inst{11}
% ...
}
\institute{
Department of Physics, Tokai University, 4-1-1, Kita-Kaname, Hiratsuka, Kanagawa 259-1292, Japan
\and Institute for Cosmic Ray Research, University of Tokyo, 5-1-5, Kashiwa-no-ha, Kashiwa, Chiba 277-8582, Japan
\and INFN and Università degli Studi di Siena, Dipartimento di Scienze Fisiche, della Terra e dell'Ambiente (DSFTA), Sezione di Fisica, Via Roma 56, 53100 Siena, Italy
\and Université Paris-Saclay, Université Paris Cité, CEA, CNRS, AIM, F-91191 Gif-sur-Yvette Cedex, France
\and FSLAC IRL 2009, CNRS/IAC, La Laguna, Tenerife, Spain
\and Departament de Física Quàntica i Astrofísica, Institut de Ciències del Cosmos, Universitat de Barcelona, IEEC-UB, Martí i Franquès, 1, 08028, Barcelona, Spain
\and Instituto de Astrofísica de Andalucía-CSIC, Glorieta de la Astronomía s/n, 18008, Granada, Spain
\and IPARCOS-UCM, Instituto de Física de Partículas y del Cosmos, and EMFTEL Department, Universidad Complutense de Madrid, Plaza de Ciencias, 1. Ciudad Universitaria, 28040 Madrid, Spain
\and INAF - Osservatorio Astronomico di Roma, Via di Frascati 33, 00040, Monteporzio Catone, Italy
\and INFN Sezione di Napoli, Via Cintia, ed. G, 80126 Napoli, Italy
\and Max-Planck-Institut für Physik, Föhringer Ring 6, 80805 München, Germany
% ...
}

%%%%%%%%%%%%%%%%%%%%%%%%%%%%%%%%%%%%%%%%%%%%%%%%%%%%%%%%%%%%%%%%%%%%%%%%%%%%%%%%%
%
% ABSTRACT
%
% \abstract{}{}{}{}{} 
% 5 {} token are mandatory
\abstract
{
    There are currently three pulsars firmly detected %above 20 GeV 
    by imaging atmospheric Cherenkov telescopes (IACTs), two of them reaching the TeV energy range, challenging %standard models of high-energy emission of pulsars.
    models of very-high-energy (VHE) emission in pulsars.
    %The study of gamma-ray pulsars above tens of GeV is required to study whether very-high-energy emission is common in these objects and to model it. Therefore, new instruments are needed to overcome the current sensitivity of IACTs. 
    More precise observations are needed to better characterize pulsar emission at these energies.
    The LST-1 is the prototype of the Large-Sized Telescope, that will be part of the Cherenkov Telescope Array Observatory (CTAO). Its improved performance over previous IACTs makes it well suited for studying pulsars. 
}
{
    In this work we aim to study the Crab pulsar emission with the LST-1, improving and complementing the results from other telescopes. 
    %In addition, the Crab pulsar observations can be used to characterize the potential of the LST-1 to detect and study other pulsars.
    Crab pulsar observations can also be used to characterize the potential of the LST-1 to study other pulsars and detect new ones.
}
{
    We analyzed a total of $\sim$103 hours of gamma-ray observations of the Crab pulsar conducted with the LST-1 in the period from September 2020 to January 2023. % and performed at zenith angles less than 50 degrees. 
    The observations were carried out at zenith angles less than 50 degrees.    
    %A new analysis of {\it Fermi}-LAT data, including data from $\sim$14 years, was also performed in order to do joint studies with the LST-1 data.
    To characterize the Crab pulsar emission over a broader energy range, a new analysis of the Fermi-LAT data was also performed, including $\sim$14 years of observations.
    }
{
    %The Crab Pulsar phaseogram, long-term light-curve, and phase-resolved spectra  were reconstructed from 20 GeV up to 700 GeV, overlapping in the {\it Fermi}-LAT energy range with high statistical significance.  The pulsed emission was detected at a significance level of $\sim$15$\sigma$ and the power law extension at $E>$500 GeV is confirmed. We found spectral indices of $\sim$3.44 and $\sim$3.03 for Peak 1 and Peak 2, respectively. The bridge emission was also detected significantly. A joint {\it Fermi}-LAT and LST-1 analysis was performed, showing a smooth transition between both instruments. The high significance of the detection of the first pulsar with the LST-1 points to an excellent LST performance for this kind of source.
    %
    The Crab pulsar phaseogram, long-term light-curve, and phase-resolved spectra is reconstructed with the LST-1 from 20 GeV to %700 GeV.
    450 GeV for P1 and up to 700 GeV for P2.
    The pulsed emission is detected with a significance level of $15.2\sigma$.
    The two characteristic emission peaks of the Crab pulsar are clearly detected ($>$10$\sigma$), as well as the so-called bridge emission between them ($5.7\sigma$). 
    %The first peak ($10.5\sigma$) extends up to an energy of XX GeV and the second peak ($12.1\sigma$) up to XX GeV. 
    We find that both peaks are well described by power laws, with spectral indices of $\sim$3.44 and $\sim$3.03 respectively.  
    The joint analysis of {\it Fermi}-LAT and LST-1 data shows a good agreement between both instruments in the overlapping energy range. %, allowing the Crab pulsar emission to be characterized with high statistical significance.
    The detailed results obtained in the first observations of the Crab pulsar with LST-1 show the potential that CTAO will have to study this type of sources.    
}
{}


\maketitle

%%%%%%%%%%%%%%%%%%%%%%%%%%%%%%%%%%%%%%%%%%%%%%%%%%%%%%%%%%%%%%%%%%%%%%%%%%%%%%%%%
% 
%  INTRO
%
\section{Introduction}
Pulsars are highly magnetized and rapidly rotating neutron stars (NS) that emit beamed radiation from radio up to gamma rays. Although almost 300 gamma-ray pulsars have been identified so far with the {\it Fermi}-LAT \citep{Fermi_LAT_3ct}, only three pulsars have been detected by imaging atmospheric Cherenkov telescopes (IACTs) at a significance level above 5$\sigma$: the Crab pulsar \citep{crab_pulsar_detection_magic, lopez2009detection}, the Vela pulsar \citep{Vela} and the Geminga pulsar \citep{Geminga}. Each of these three pulsars is unique. Two of them, Crab and Vela, have been detected at TeV energies \citep{MAGIC_teraelectronvolt,Vela_tev}. The emission and pulse profile found at such high energies cannot be easily explained with the curvature radiation models that predict suppression of the emission at a few GeV. In the case of Vela, the TeV emission is associated with a second radiation component reaching 20 TeV \citep{Vela_tev}. 

Detecting more pulsars with ground-based gamma-ray telescopes is a challenge due to their steep spectra above 10 GeV. {\it Fermi}-LAT measurements of gamma-ray pulsars show cutoffs in the spectra at a few GeVs \citep{Fermi_LAT_3ct}.  Due to the low expected fluxes from these sources above 50 GeV \citep{2015_McCann}, the current generation of IACTs is not bound to detect more pulsars. The search for more pulsars is necessary to understand whether the very-high-energy (VHE, E$>$100 GeV) emission of these objects is something unique or if there is a whole population of VHE pulsars. Therefore, improving the sensitivity of IACTs is necessary to study new gamma-ray pulsars above 10 GeV and constrain the models at those energies.

The Cherenkov Telescope Array Observatory (CTAO) \citep{Zanin:2021tx} will be the next generation of IACTs. It will be located in two sites in both hemispheres to cover the full VHE sky. CTAO will be composed of an array of multiple telescopes of different sizes, increasing over an order of magnitude the sensitivity of current IACTs. The Large-Sized telescopes (LSTs) \citep{Cortina:2019RL} will be the largest ones with a dish diameter of 23 meters, optimized for low energies (20 GeV-200 GeV). The LST-1 is a fully equipped LST prototype built at the Roque de los Muchachos observatory (ORM) on the island of La Palma, which will be part of the observatory \citep{lst_status}. It was inaugurated in 2018 and is now producing its first science results after several years of commissioning \citep{lst1_performance, lhaaso_lst1}.

The Crab pulsar and nebula are the remnants of the SN1054 event. Due to its young age, it is a very energetic pulsar ($\dot E \approx 4.6 \cdot 10^{38}$erg s$^{-1}$; \citeauthor{Lyne_2015} \citeyear{Lyne_2015}) with a rotation period of  $P\approx$ 33 ms \citep{Staelin_1968}. The Crab pulsar was first detected and studied in radio \citep{comella_pulsar}, and afterwards in almost all wavelengths. The pulsed emission from the Crab pulsar above 25 GeV was first detected by MAGIC \citep{crab_pulsar_detection_magic}, rejecting the existence of a super-exponential cutoff predicted by polar cap models \citep{Aleksi__2011}. This fact was confirmed after the study of the pulsar with {\it Fermi}-LAT, hinting at the existence of a sub-exponential cutoff in the spectrum at a few GeV \citep{Abdo_2009}, a feature that was found in other gamma-ray pulsars. The pulsar was detected in the VHE regime up to 400 GeV by MAGIC \citep{Aleksi__2012} and VERITAS \citep{veritas_2011}. A few years later, MAGIC reported the detection of the Crab pulsar up to 1.5 TeV \citep{MAGIC_teraelectronvolt}. The overall emission at VHE can be well described by a power law (PWL), supporting those models that consider inverse Compton (IC) processes in the outer magnetosphere or beyond.

In this work, we describe the analysis and results obtained from the first observations of the Crab pulsar with the LST-1. The aim is to characterize the emission of the pulsar above 20 GeV with the LST-1 and {\it Fermi}-LAT data and examine the potential of this new telescope for the study of pulsars at VHEs.

%%%%%%%%%%%%%%%%%%%%%%%%%%%%%%%%%%%%%%%%%%%%%%%%%%%%%%%%%%%%%%%%%%%%%%%%%%%%%%%%%
%
%
\section{LST-1 observations overview}

The LST-1 observed the Crab Nebula and pulsar during the first years of operation as part of its commissioning program. A total of more than 150 hours were collected from September 2020 to January 2023. The data were taken in 20-minute runs in wobble mode \citep{FOMIN1994137}, where the source is located at a 0.4-degree offset from the camera center. We applied quality cuts to the data, removing those runs with low trigger and pixel rates (see more details in \citeauthor{lst1_performance} \citeyear{lst1_performance}). In addition, we used only data taken in dark conditions.  %We also discarded those runs taken with technical problems. 
We also discarded those runs affected by technical problems.

As a result, $\sim$103 hours of observations taken at zenith distance (Zd) below 50 deg survived the quality cuts and were used in the final analysis. Out of these, 76 hours were collected at Zd $<$ 35 deg, most of them below 20 deg, decreasing the overall energy threshold down to $\sim$20 GeV in the analysis \citep{lst1_performance}. The trigger settings of the telescope were variable before August 2021, %so the threshold of the data taken before that date is less stable and slightly larger than the one collected afterwards. 
so the energy threshold of data taken before that date is less stable and slightly higher than that of data collected afterwards. 
%This improvement in the threshold is a result of the advance on the commissioning of the LST-1.
The improvements in the telescope threshold are a consequence of the advances made during the commissioning of LST-1.
 
%%%%%%%%%%%%%%%%%%%%%%%%%%%%%%%%%%%%%%%%%%%%%%%%%%%%%%%%%%%%%%%%%%%%%%%%%%%%%%%%%
%
%
\section{Data analysis}

%--------------------------------------------------------------------------
%
%
\subsection{LST-1 data analysis}

LST-1 data were reduced using \texttt{cta-lstchain} v0.9.14 \citep{lstchain_proc,lstchain_zenodo}, a software designed for the data analysis of the LST-1, following the usual IACT analysis chain. This allowed us to clean and parametrize the images produced in the camera by atmospheric showers. The image parametrization is used to infer the direction and energy (called reconstructed energy) of the primary particle through trained Random Forest (RF) algorithms \citep{Albert_2008}. An additional parameter called \textit{gammaness} is computed, defined as a score that rises for the higher resemblance of the event to a gamma-ray initiated one. To optimize the analysis of faint showers (i.e. low-intensity images), we included in the training some parameters that depend on the known position of the source in the camera plane in the so-called source-dependent approach. This improves the performance with respect to the standard source-independent analysis (see Sect. 4.1.4). One of the source-dependent parameters added is \textit{alpha}, defined as the angle between the major axis of the fitted shower ellipse and the line that joins the center of gravity of the image and the assumed source position in the camera plane.
   
We used Monte Carlo (MC) simulations to evaluate the performance of the telescope. The MC simulations of gamma-ray-initiated showers used in this work are part of an all-sky MC production simulated in declination lines \citep{lst1_performance}. The one used to analyze the data sample is the closest to that of the Crab pulsar (22.76 deg). The MC data were tuned by adding Poissonian noise to match the real night sky background of the Crab pulsar region. The MC sample was processed with the help of the \texttt{lstmcpipe} package \citep{lstmcpipe, lstmcpipe_zenodo}. Two samples were produced: a sample to train the RF, and a test sample to characterize the response of the telescope. The MC test dataset was simulated on a grid of nodes with different zenith/azimuth pointings in the sky. To calculate the instrument response functions (IRFs), we used the nodes closest to the pointing of the telescope during the observations. This way, it is possible to account for the dependence of the telescope performance on the airmass and the angle formed by the orthogonal component of the geomagnetic field and the pointing of the telescope. 

After the reconstruction of the events, several cuts were applied. First, an intensity %cutoff
cut is needed to provide a common analysis threshold for the entire data sample and a good match between observed data and MC, as explained in \citet{lst1_performance}.  We applied an overall intensity cut of 80 photoelectrons (p.e.) for all the data taken before August 2021, and an intensity cut of 50 p.e. to the data taken after that date to account for the different trigger thresholds of the telescope during these periods (see Sect. 2). In addition, we applied energy-dependent cuts on the direction (\textit{alpha}) and \textit{gammaness}, computed by setting a 70\% MC efficiency on the gamma MC sample for each of the cuts separately. 

The last step to analyze the pulsar is to obtain the phase of the rotation of the star associated with each event. For that, we used the \texttt{PINT} package v0.9.3 \citep{pint} and the Crab pulsar ephemeris provided by the Jodrell Bank observatory \citep{Lyne_1993}, available in the web address \url{http://www.jb.man.ac.uk/~pulsar/crab.html}. Finally, the spectral results of the analysis were produced with \texttt{Gammapy} v1.0.1 \citep{gammapy_2023, gammapy_zenodo}. As a consistency check for the analysis, we compared (a posteriori) the weighted distributions of the MC shower parameters with those of the pulsed excess, finding a good agreement.

The Crab pulsar is characterized by showing two emission peaks in each rotation, which remain aligned at all wavelengths. The first peak, located at phase 0, is defined as P1 and is the most intense in radio and also in the {\it Fermi}-LAT sample between 100 MeV and 1 GeV. The second peak, P2, is however the most intense at VHEs. For the analysis, we adopted the phase intervals defined in \citet{Aleksi__2012}, namely P1=[-0.017, 0.026] and P2=[0.377, 0.422]. The background level was estimated using the OFF region [0.52, 0.87], %where no hint of pulsed emission is present. 
where no pulsed emission is expected.

%--------------------------------------------------------------------------
%
%
\subsection{{\it Fermi}-LAT  data analysis}

Since its launch in 2008, the {\it Fermi} Large Area Telescope ({\it Fermi}-LAT) has been observing the gamma-ray sky continuously in the energy range between 20 MeV to hundreds of GeV. 
To study the Crab pulsar emission at energies lower than those accessible to the LST-1, we analyzed public {\it Fermi}-LAT data taken from the 4th of August 2008 until the 24th of August 2022. 
%To achieve a larger energy coverage in our study we analyzed public {\it Fermi}-LAT data taken between 100 MeV and 100 GeV from the 4th of August 2008 until the 24th of August 2022. 
This resulted in $\sim$14 years of observations, extending the sample used in previous works \citep{Yeung_2020}. 
%With the {\it Fermi}-LAT sample, we can study the Crab pulsar emission at energies lower than those accessible to the LST-1. 
 
We processed this data set using the {\it Fermi} Science Tools version v11r5p3 \citep{fermitools} and the P8R2\_SOURCE\_V6 instrument response functions (IRFs). 
We selected events classified as event class 128 (`Source') and event type 3 from a circular region of interest (ROI) of 15 deg centered at the Crab pulsar coordinates, RA=$05^{\rm h}34^{\rm m}31.9^{\rm s}$, Dec=$22^\circ 00^{'} 52.2^{''}$. 
To reject the background coming from the Earth’s limb, we excluded time intervals where the ROI was observed at zenith angles greater than 90 deg.
The pulsar rotational phases were computed using the {\it Tempo2} package \citep{tempo2} with the same ephemeris as for the LST-1 data analysis. Phase-filtered event files were produced, containing only photons in the OFF phase region, %where no pulsar emission is expected, 
or in the phase region corresponding to each of the pulsar emission peaks defined above.

For the spectral reconstruction, we performed a binned likelihood analysis using the {\it pyLikelihood} python module of the {\it Fermi} Science Tools, with a bin size of $0.2^\circ$ per pixel and 40 logarithmically spaced energy bins between 100 MeV and 2 TeV. 
The initial spectral-spatial model included all sources from the LAT 10-year source catalog (4FGL) \citep{4FGL} within the ROI that was expanded by 5 deg to account for partially contained sources. 
The spectral parameters for sources with a significance higher than $5\sigma$ and located within 5 deg of the center of the ROI were left free. Also, the normalization factor of the Galactic (gll\_iem\_v07.fits) and isotropic background (iso\_P8R3\_SOURCE\_V3\_v1.txt) models were let free.
For the rest of the sources, the spectral parameters were set to their catalog values.
After the first fit, all sources with $TS<4$ were removed from the model.
We then use the events in the OFF region to characterize the gamma-ray background due to emission from the Crab Nebula, whose IC and Synchrotron components appear as two different sources in the catalog, J0534.5+2201i and 4FGL J0534.5+2201s respectively. After this, the Crab Nebula spectral parameters were left fixed, scaling only the normalization factor to account for the different phase widths of the off-pulse and peak regions.
Finally, the spectra of P1 and P2 were analyzed independently, using smooth broken power law models.
%The Crab pulsar model used for the fit was a smooth broken power law. 
To obtain the spectral points we repeated the spectral fit in each energy bin using a power-law model with a fixed spectral index of $2$ and with the normalization factor free. Only spectral points with a significance higher than $2\sigma$ are shown in the plots.

%%%%%%%%%%%%%%%%%%%%%%%%%%%%%%%%%%%%%%%%%%%%%%%%%%%%%%%%%%%%%%%%%%%%%%%%%%%%%%%%%

%%%%%%%%%%%%%%%%%%%%%%%%%%%%%%%%%%%%%%%%%%%%%%%%%%%%%%%%%%%%%%%%%%%%%%%%%%%%%%%%%
%
%
\section{Results}

%--------------------------------------------------------------------------
%
%
\subsection{Phaseogram}

\subsubsection{Pulsed signal}
  
\begin{figure*}[!ht!]
    \resizebox{\hsize}{!}{\includegraphics{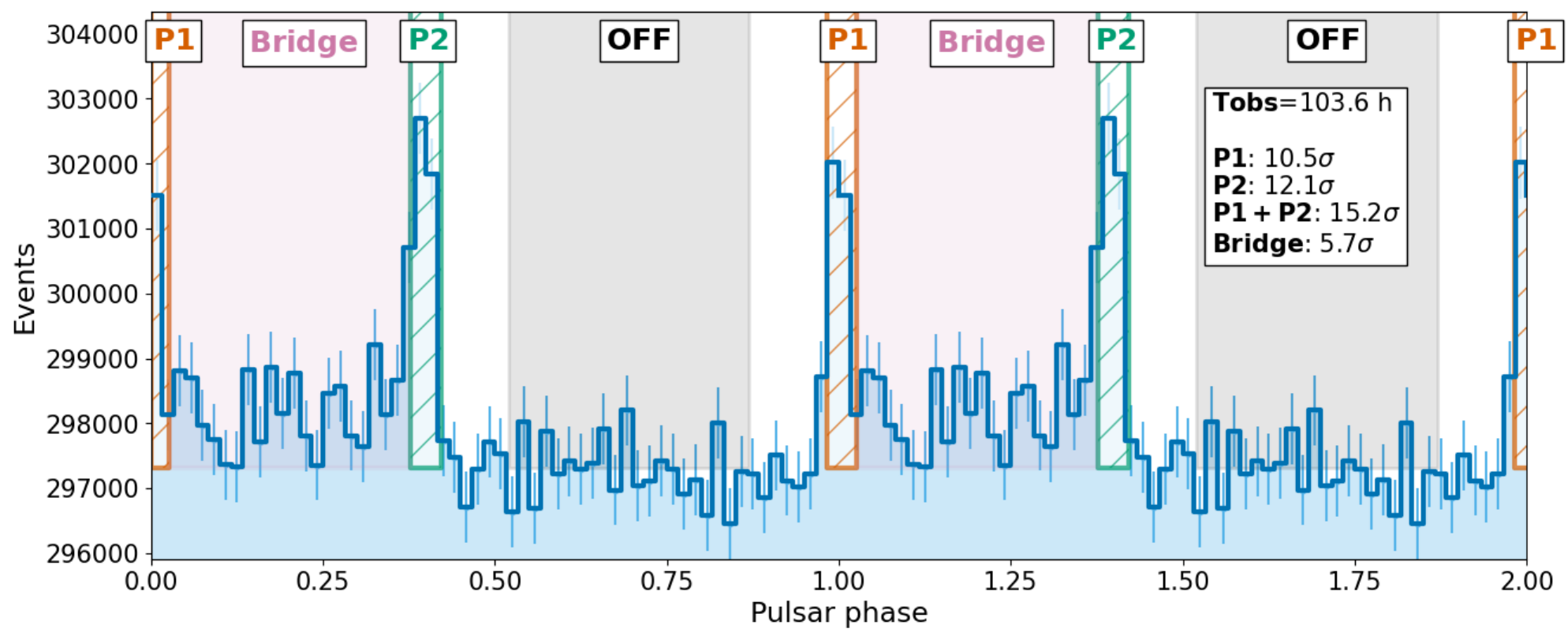}}
    \caption{Phaseogram of the Crab pulsar sample from LST-1 data. Both peaks and the overall bridge emission between peaks are detected significantly. The energy threshold in the sample (Zd$<$50 deg) is E$_{\rm{th}}$$\sim$ 20 GeV.  The period corresponding to two rotations is shown in the phaseogram for a better visualization.}
    \label{phaseogram}
\end{figure*}

The phase-folded phaseogram obtained with the LST-1 is shown in Fig.~\ref{phaseogram}. P1 and P2 are detected at a statistical significance of 10.5$\sigma$ and 12.1$\sigma$ respectively, computed using formula (17) in \citet{Lima_1983}. The joint pulsed emission (P1+P2) is detected at 15.2$\sigma$. Between both peaks a fainter signal, commonly known as the bridge emission, is found. In this work, we use the two definitions for the bridge used in \citet{crabpulsar_magic_2014}. Defining the bridge as the whole region between peaks (i.e. Bridge$_{M}$ = [0.026,0.377]) the emission is detected at a significance level of 5.7$\sigma$. If we redefine the bridge region as done in \citet{fierro_1998} (Bridge$_{E}$ = [0.14,0.25]) the significance is 3.7$\sigma$.

\begin{figure}[!t]
    \centering
    \includegraphics[width=\linewidth]{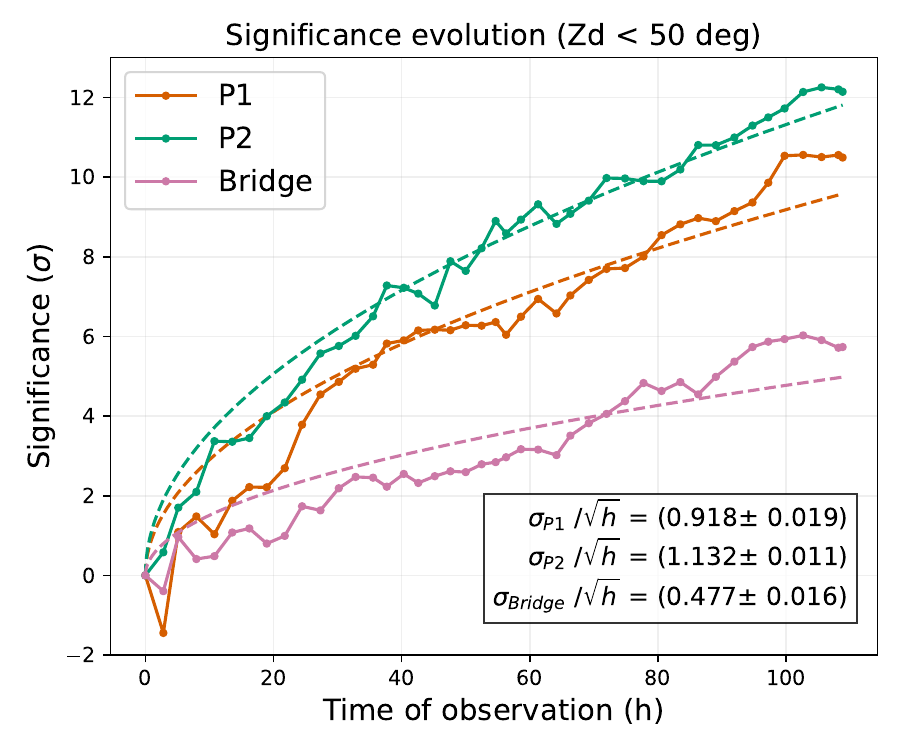}   
    \caption{Evolution of the significance  with the total time of observation for P1, P2, and the bridge emission, defined as Bridge$_{M}$ = [0.026,0.377].}
    \label{sig_vs_time}
\end{figure}

The increase of the signal with time, as seen in Fig.~\ref{sig_vs_time}, confirms the stability of the analyzed data sample. We did a fit of the data finding that the evolution can be characterized by $\sigma_{\rm P1}$(h)=(0.916 $\pm$ 0.021)h$^{1/2}$, $\sigma_{\rm P2}$(h)=(1.133 $\pm$ 0.012)h$^{1/2}$ and $\sigma_{\rm Bridge}$(h)=(0.478 $\pm$ 0.018)h$^{1/2}$, where $h$ is the total number of hours of observation. These values change if we limit our sample to lower zenith angles. For instance, at Zd<35 deg, the values increase up to $\sigma_{\rm P1}$(h)=(1.109 $\pm$ 0.016)h$^{1/2}$, $\sigma_{\rm P2}$(h)=(1.272 $\pm$ 0.014)h$^{1/2}$ and $\sigma_{\rm Bridge}$(h)=(0.619 $\pm$ 0.017)h$^{1/2}$.  These results highlight the good performance of the LST-1. For comparison, the stereo MAGIC SumTrigger-II reported an overall detection rate of $\sigma_{\rm P1+P2}$ =2.0h$^{1/2}$ for the Crab pulsar at Zd$<$25 deg \citep{Ceribella_crabpulsar}, similar to the detection rate of a single LST-1 telescope ($\sigma_{\rm P1+P2}$ $\approx$ 1.8h$^{1/2}$) at the same zenith.

%--------------------------------------------------------------------------
%
%
\subsubsection{Morphology of the peaks}

\begin{figure*}[!t]
    \resizebox{\hsize}{!}{\includegraphics{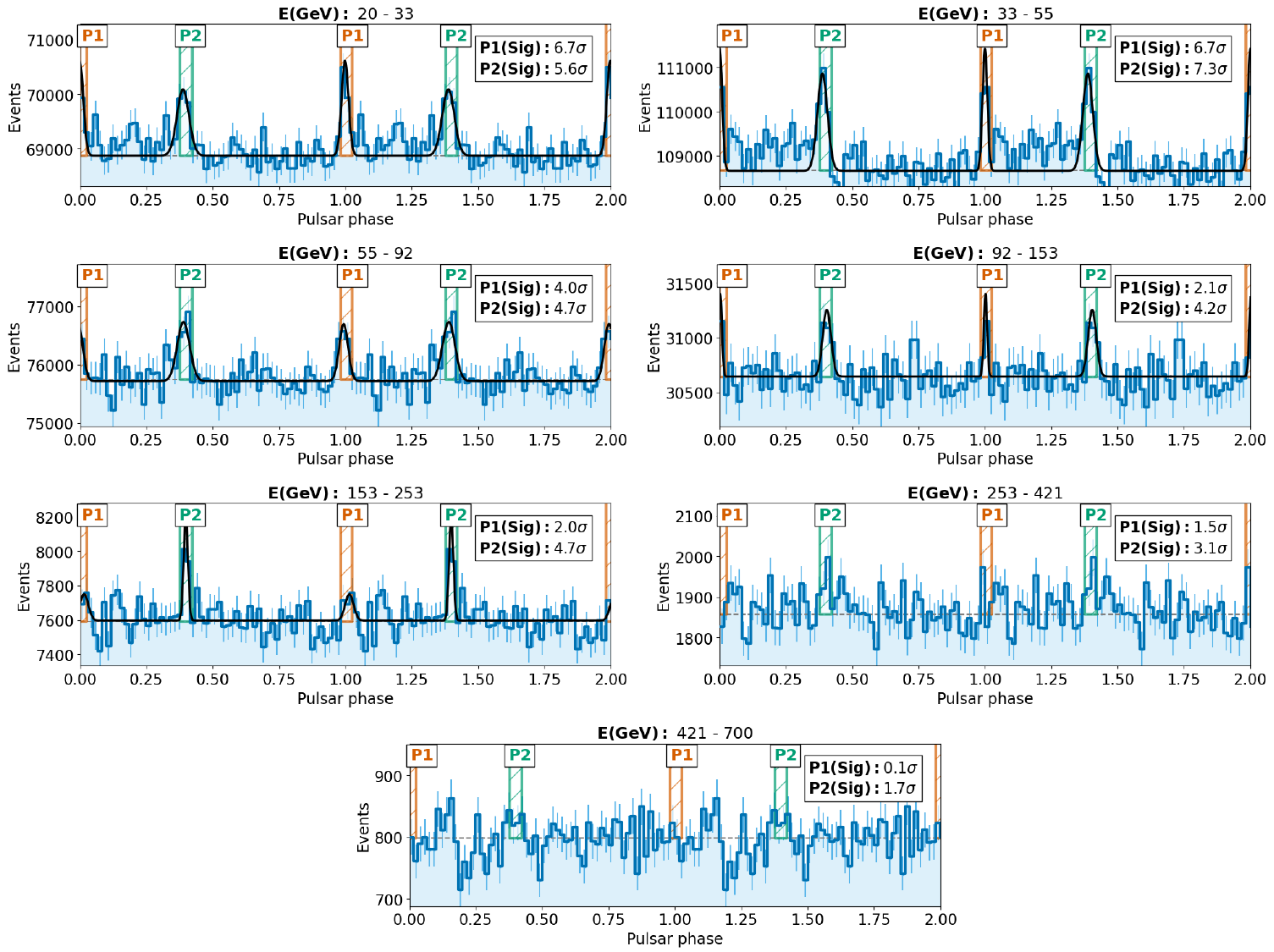}}
    \caption{Phaseogram of the Crab pulsar from LST-1 data in different energy bins from 20 GeV to 700 GeV. The statistical significance of each peak is given in each plot. The black line shows the best fits to the pulse profile. Above 250 GeV the fit was not successful since the signal of P1 begins to disappear. }
    \label{phaseogram_energy}
\end{figure*}

\begin{figure}[!htp]
    \centering
    \includegraphics[width=\hsize]{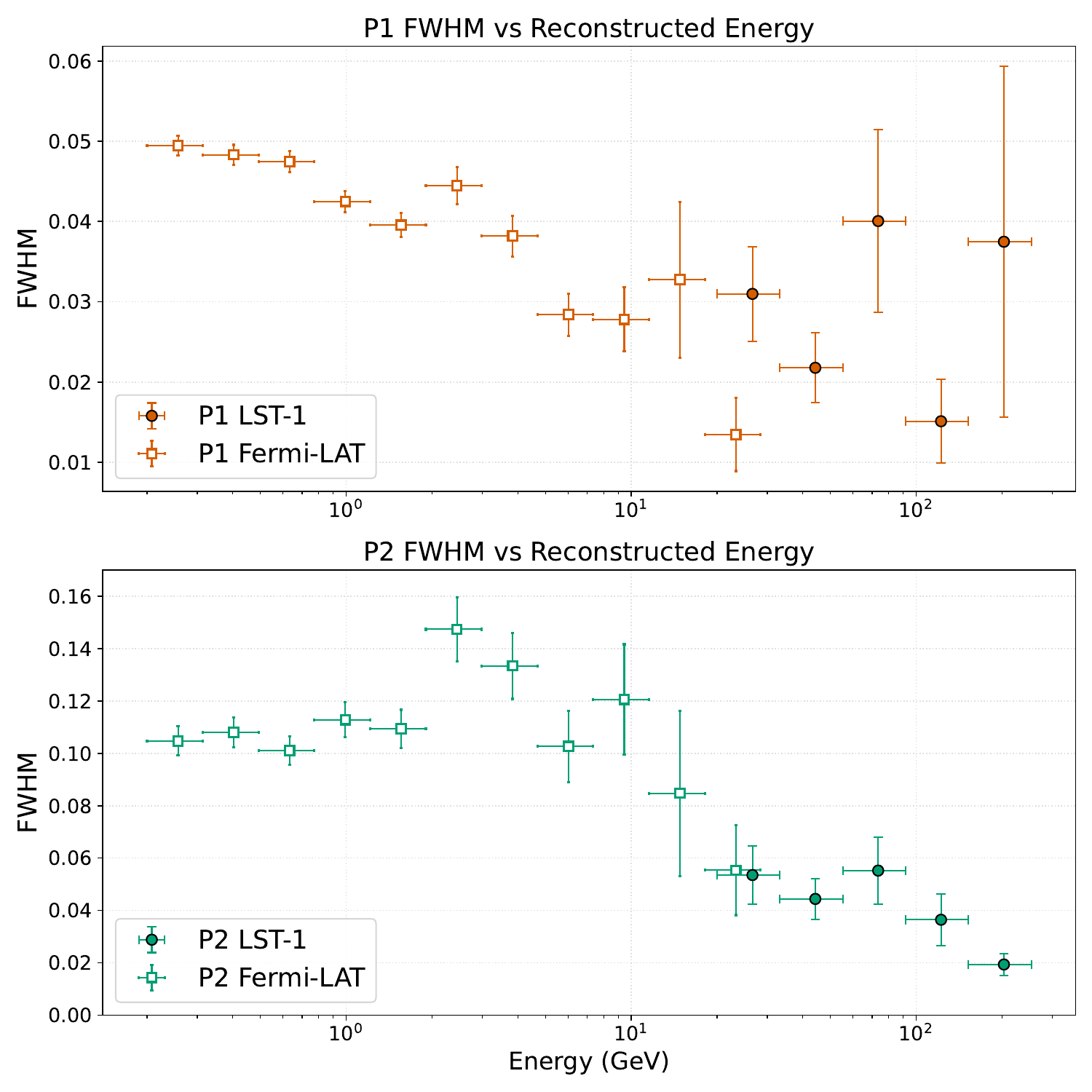}  
    \caption{Evolution of the peak width as a function of the energy from 100 MeV to 200 GeV using {\it Fermi}-LAT and LST-1 data. The fit of the LST-1 data was not successful above 200 GeV due to the lack of statistics.}
    \label{fwhm_energy}
\end{figure}

The phaseogram was also studied in different energy bins. In particular, we divided our sample into seven bins from 20 GeV to 700 GeV (see Fig.~\ref{phaseogram_energy}), approximately 5 bins per decade. %in the energy region where signal of either of each peaks is found. 
Assuming that the peaks follow symmetric Gaussian distributions, we fitted the phaseogram to a double Gaussian model with an overall background to study the morphology of the peaks. The values of the mean phase and width of each peak are shown in Table \ref{table:fwhm}. The peak positions do not shift significantly. The width of P2 seems to decrease with energy (see Fig.~\ref{fwhm_energy}). This feature was already found in other studies \citep{Aleksi__2012}, being crucial to understand emission models at energies greater than 100 GeV \citep{Harding_2021}. The LST-1 measurement in Fig.~\ref{fwhm_energy} was fitted to a linear model (FWHM = m$\cdot$log(E)+n) above 20 GeV, finding that for P2 the best fit has a slope of m$_{P2} = 0.041 \pm 0.009$ and shows a $p_{\rm value}$ = 0.65. For P1 the fitted model to the LST-1 data shows a slope of m$_{P1} = 0.016 \pm 0.013$. Although for this model $p_{\rm value}$ = 0.31, the large statistical uncertainties of the LST-1 points %and the low slope found 
make it difficult to conclude a significant variation of the width of P1 above 20 GeV. % To test the possibility of an energy-independent width for P1 above that energy, the LST-1 data was fit to a constant value, resulting in an average value of FWHM$_{P1}$ = 0.023 $\pm$ 0.004 with a $p_{\rm value}$ = 0.14.  

\begin{table*}[h]
    \caption{Peak position ($\mu$) and width (FWHM) of each peak, P1 and P2, after fitting the phaseogram in each energy bin to a double Gaussian model. }             % title of Table
    \label{table:fwhm}      % is used to refer this table in the text
    \centering                          % used for centering table
    \begin{tabular}{c c c c c}        % centered columns (4 columns)
    \hline\hline                 % inserts double horizontal lines
    Energy (GeV)  & $\mu_{1}$ & FWHM$_{1}$ ($\cdot 10^{-2}$) & $\mu_{2}$ & FWHM$_{2}$ ($\cdot 10^{-2}$) \\    % table heading 
    \hline                        % inserts single horizontal line
    20 - 33 & 0.999 $\pm$ 0.003 & 3.1 $\pm$ 0.6  & 0.389 $\pm$ 0.004 & 5.4 $\pm$ 1.1 \\
    33 - 55  & 1.0000 $\pm$ 0.0018 & 2.2 $\pm$ 0.4  & 0.387 $\pm$ 0.003 & 4.4 $\pm$ 0.8 \\ 
    55 - 92 & 0.994 $\pm$ 0.005 & 4.0 $\pm$ 1.1  & 0.388 $\pm$ 0.006 & 5.5 $\pm$ 1.3 \\ 
    92 - 153   & 1.0020 $\pm$ 0.0022 & 1.5 $\pm$  0.5 & 0.402 $\pm$ 0.004 & 3.6 $\pm$ 1.0 \\ 
    153 - 253   & 1.015 $\pm$ 0.009 & 3.7 $\pm$ 2.2  & 0.3981 $\pm$ 0.0017 & 1.9 $\pm$ 0.7 \\ 
    \hline                                   %inserts single line
    \end{tabular}
\end{table*}
    
The {\it Fermi}-LAT data were also divided into energy bins and the phaseogram was fitted to the same model as for the LST-1 data. Representing the width of the peaks as a function of energy from MeV to GeV (Fig.~\ref{fwhm_energy}) one can see a soft transition between {\it Fermi}-LAT and LST-1 data. For both peaks, the width above 20 GeV is lower than at 200 MeV as seen in other works \citep{veritas_2011, Aleksi__2012}. {\it Fermi}-LAT and LST-1 results in Fig.~\ref{phaseogram_energy} are compatible in their overlapping energy region for P2. For P1, the FWHM points of both instruments are compatible if we add a systematic error between both instruments of at least $\sim$20\%, although the statistical uncertainties dominate over the systematic ones in this case.

%--------------------------------------------------------------------------
%
%
\subsubsection{P1/P2 ratio}
\begin{figure}[t]
    \centering
    \includegraphics[width=\hsize]{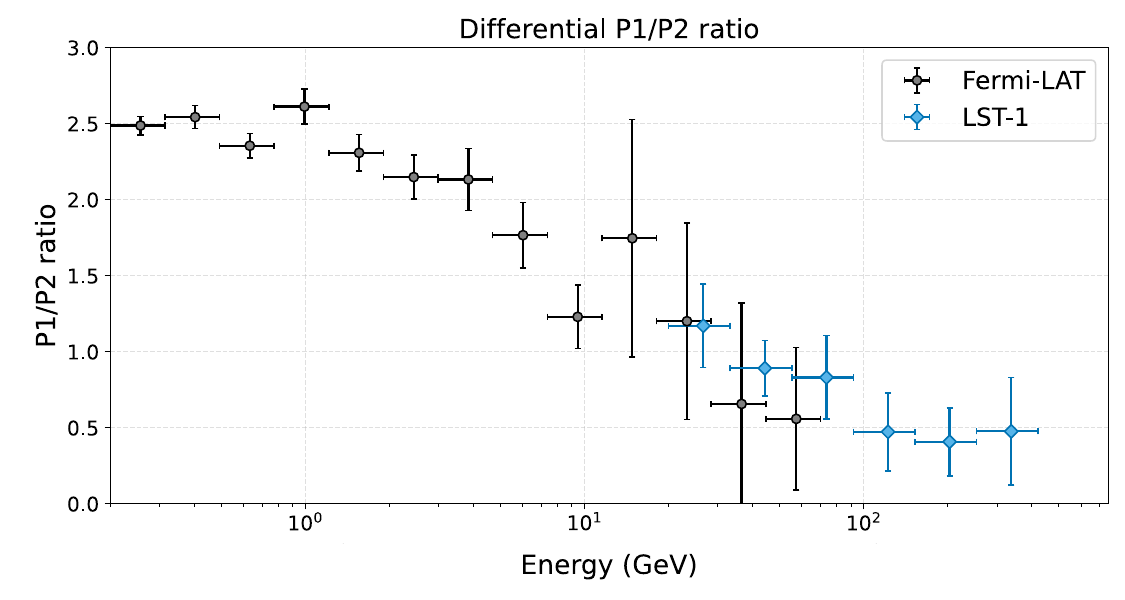}  
    \caption{Evolution of the P1/P2 ratio as a function of the energy from 100 MeV to 400 GeV using {\it Fermi}-LAT and LST-1 data. The fit of the LST-1 data was not successful above 400 GeV due to the lack of statistics for P1. }
    \label{p1p2_ratio}
    \end{figure}

As seen in Fig.~\ref{phaseogram_energy}, the intensity and significance of P1 is higher in the lowest energy bin, below 30 GeV. In the rest of the bins, P2 appears stronger than P1. To study this trend, the LST-1 differential ratio of P1/P2 was determined as well from the excess counts in each reconstructed energy bin. The same ratio was computed with the {\it Fermi}-LAT sample in 13 energy bins to plot the energy evolution of the differential ratio. As a result, we covered the energy range from 100 MeV up to 400 GeV using both instruments. The result is depicted in Fig.~\ref{p1p2_ratio}.  One can see a fast decrease of the ratio down to $\sim$0.5 at $\sim$200 GeV. This trend was already reported in other works \citep{p1p2_ratio_magic}. The P1/P2 ratio achieves 1 at E$_{\rm eq}\approx$30 GeV. The overall LST-1 ratio, integrated over the entire energy range, is P1/P2 = 0.84 $\pm$ 0.11. The LST-1 points show lower statistical errors than the Fermi-LAT ones, indicating that the LST-1 can provide more accurate results above 20 GeV even with only 100 hours.  %The integral P1/P2 ratio, computed by integrating all the excess counts in each signal region above a certain energy, was also calculated. In this case, a finer binning was used to reduce the uncertainties. The results of the evolution of the integral ratio with energy are shown in the second panel of Fig.~\ref{p1p2_ratio}. %

\begin{figure*}[!htp]
    \resizebox{\hsize}{!}{\includegraphics{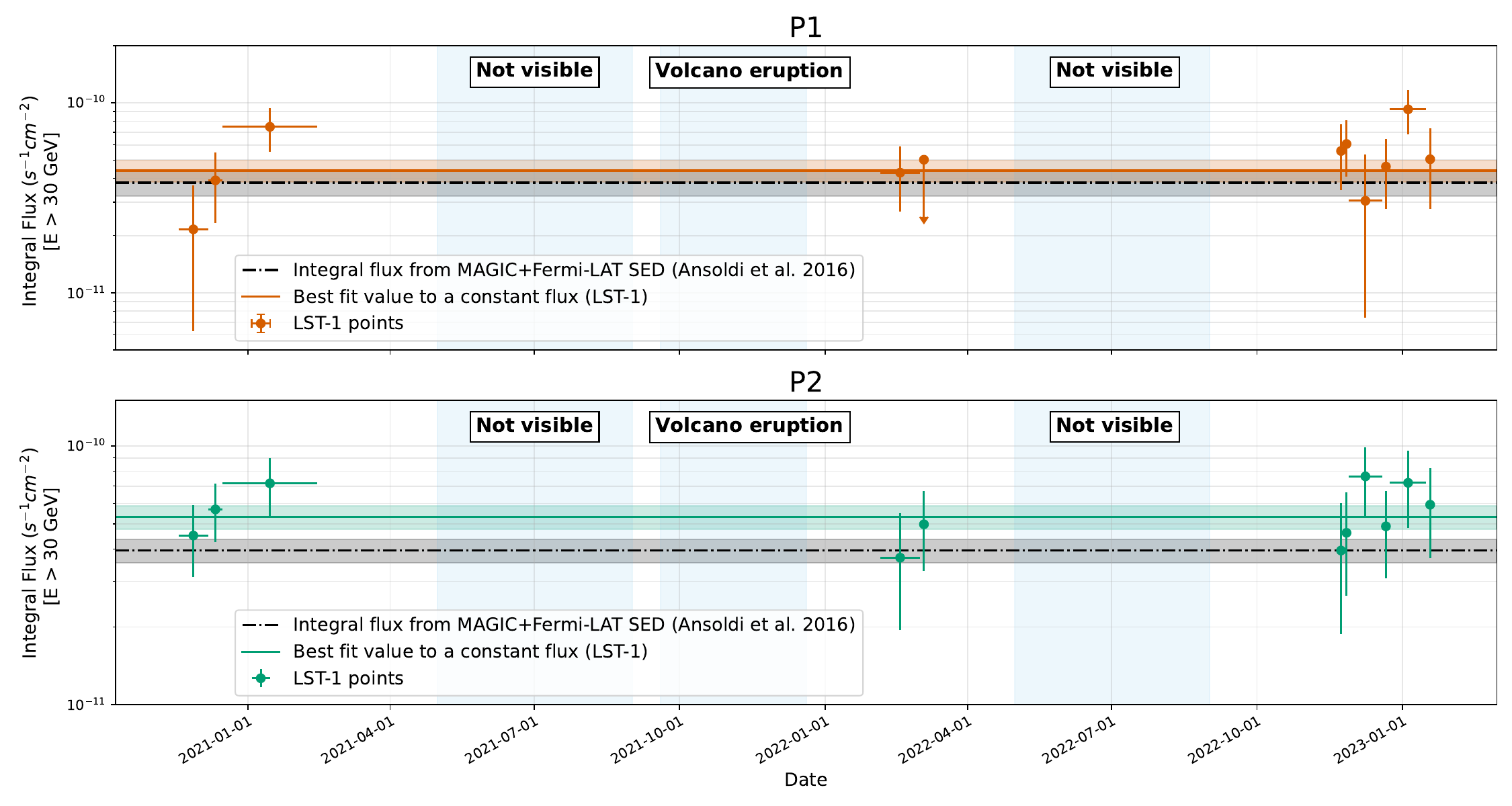}}
    \caption{Long term light-curve of P1 and P2 Crab emission above 30 GeV. Each variable time bin contains 1500 excess events in the combined phase regions P1 and P2. This value was chosen to reach at least $\sim$3$\sigma$ for the entire pulsed emission (P1+P2) in each bin. The horizontal bars indicate the time range of each bin. The flux was fit to a constant function, shown by the green line. The green dashed area represents the statistical uncertainties of the fitted flux. The reference integrated flux above 30 GeV using the MAGIC+{\it Fermi}-LAT SED reported in \citet{MAGIC_teraelectronvolt} was included in gray. The regions where the Crab pulsar was not observable are shown in blue, namely two summer periods and the volcano eruption that took place from September to December 2021 in La Palma.}
    \label{lightcurve}
\end{figure*}
   
The P1/P2 ratio points of LST-1 derived in Fig.~\ref{p1p2_ratio} are represented in reconstructed energy. Near the threshold of the LST-1 the reconstructed energy of the events is systematically greater than the true one. The maximum systematic error in the differential P1/P2 computation at low energies due to the energy dispersion of our system is $\sim$ 20\% as estimated from a set of MC simulations with a similar zenith distribution as our data. For the integral ratio, the maximum of this systematic error drops to $\sim$ 12\%. Thus, the the LST-1 P1/P2 ratios in each reconstructed energy container are therefore overestimated with respect to those of {\it Fermi}-LAT by at most that 20\%.

%--------------------------------------------------------------------------
%
%
\subsubsection{Source-dependent vs source-independent}
  
The phaseogram was computed using the source-independent analysis as well (i.e. not including source-dependent parameters in the RF training). 
%This is the standard approach in IACT data analysis. 
The analysis chain was similar to the one used for the source-dependent case but changing the MC efficiency to 91\% to have a similar background rate in both approaches. In particular, for this efficiency, we get a difference in background level $< 1\%$. The results are shown in Table \ref{table:src_indep}. The source-dependent analysis shows better performance for studying the pulsed emission, with a difference of 1.5$\sigma$ in P1 and 2.7$\sigma$ in P2. \\

The results described in \citet{lst1_performance} show that the sensitivity curves below 100 GeV are similar for both source-dependent and source-independent analysis. The difference found in the Crab pulsar analysis indicates that the source-dependent approach improves the sensitivity at the lowest true energies, near the threshold of the telescope, where the signal of the pulsar is more intense and the background estimation in the Crab Nebula is more uncertain.

\begin{table}
    \caption{Comparison of the pulsed signal using source-dependent and source-independent approaches in the RF training}
    \centering
    \begin{tabular}{ccccc}
        \hline\hline
        $\textbf{Type}$ & P1 + P2  & P1   & P2 & Bridge  \\
        \hline
        Source-dependent   & 15.2 $\sigma$   & 10.5 $\sigma$              & 12.1 $\sigma$ & 5.7 $\sigma$         \\
        \hline
        Source-independent   & 12.5  $\sigma$   & 9.2 $\sigma$              & 9.5 $\sigma$ & 4.4 $\sigma$ \\
        \hline
    \end{tabular}
    \label{table:src_indep}
\end{table}

%--------------------------------------------------------------------------
%
%
\subsection{Long-term light-curve}

In order to test for possible variability in the Crab pulsar emission, we computed the long-term light-curve of the pulsed emission for both P1 and P2 above 30 GeV. The minimum energy was selected to minimize the effect of energy migration near the telescope threshold for the variability studies. We divided the sample into variable time bins requiring a fixed number of excess events for the total pulsed emission. We selected a value of N$_{\rm ex}$=1500 to achieve at least $\sim$3$\sigma$ for the entire pulsed emission in each bin. In Fig.~\ref{lightcurve} the corresponding long-term light-curve for each peak is shown. 

We fitted the flux points to a constant function and performed a $\chi^{2}$ test on the data. We found a value of $\chi^2/\textrm{ndf}=13.1/$10 ($p_{\rm value}=0.22$) and $\chi^2/\textrm{ndf}=4.7/$10 ($p_{\rm value}=0.91$) for P1 and P2, respectively. For the total pulsed emission, a value of $\chi^2/\textrm{ndf}=12.8/$10 ($p_{\rm value}=0.24$) is found. Therefore, no hint of variability is detected in the sample, and the
fluxes are compatible with constant functions of f$_{\rm P1}$=(4.4 $\pm$ 0.6)$\cdot$10$^{-11}$cm$^{-2}$s$^{-1}$ and f$_{\rm P2}$=(5.3 $\pm$ 0.5)$\cdot$10$^{-11}$cm$^{-2}$s$^{-1}$  showing a weighted relative root mean squared error of 46\% for P1 and 22\% for P2.  For comparison, we calculated the integral flux using the joint MAGIC and {\it Fermi}-LAT Spectral Energy Distribution (SED) reported in \citet{MAGIC_teraelectronvolt} above 30 GeV. For P1, we obtained a value of f$_{\rm ref, P1}$=(3.8 $\pm$ 0.6)$\cdot$10$^{-11}$cm$^{-2}$s$^{-1}$ compatible with the LST-1 integral flux. In the case of P2, we found a value of f$_{\rm ref, P2}$=(4.0 $\pm$ 0.4)$\cdot$10$^{-11}$cm$^{-2}$s$^{-1}$, lower than for the LST-1 sample. This could be a result of the different energy thresholds for MAGIC and for the LST-1, also reflected in the distinct spectral fits obtained for both.

%--------------------------------------------------------------------------
%
%   
\subsection{Spectral Energy Distribution of the peaks}

\begin{figure}[t]
   \centering
   \includegraphics[width=\hsize]{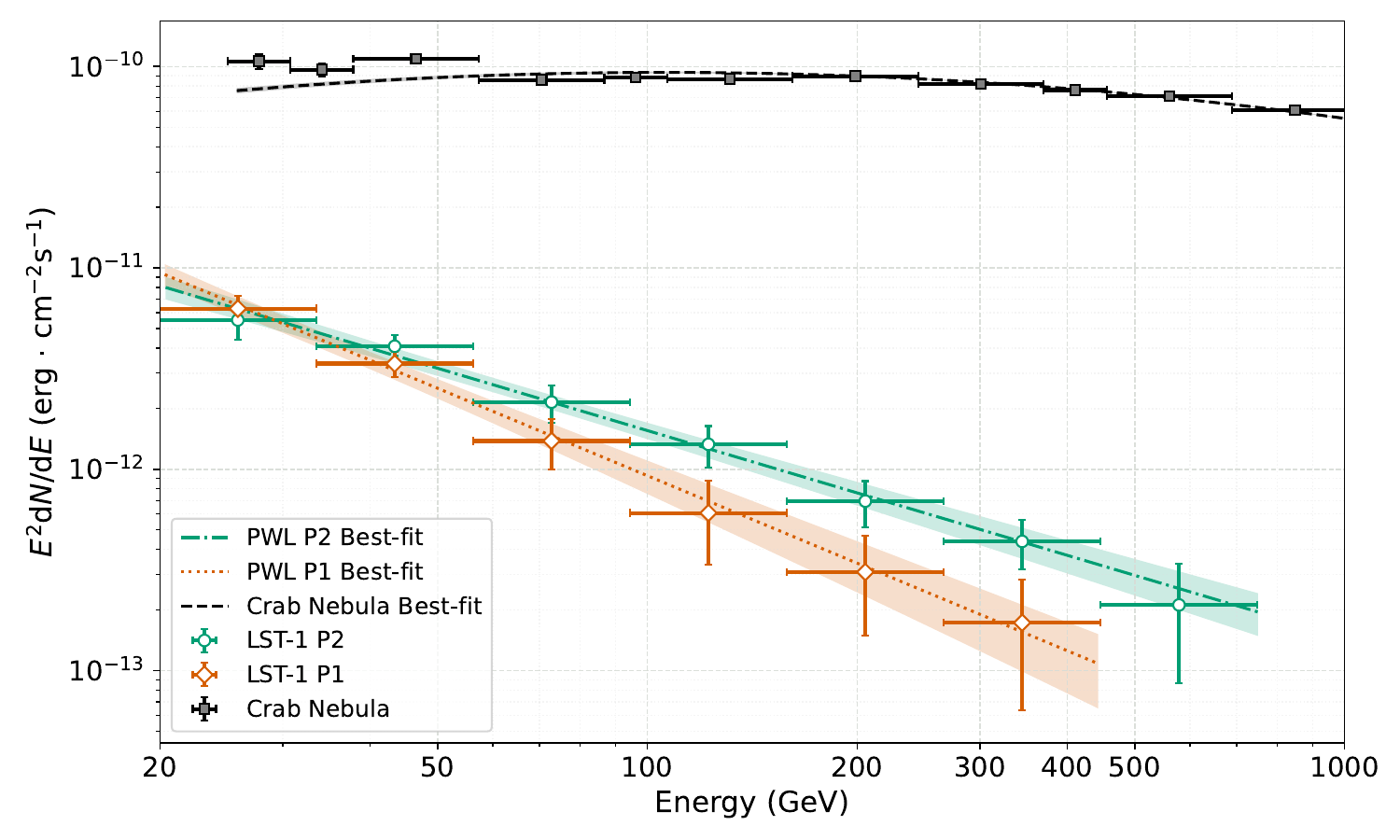}
      \caption{LST-1 SED of P1 and P2 of the Crab pulsar from 20 GeV to 700 GeV. The Crab Nebula spectrum obtained with the same sample is represented in black.}
         \label{sed}
   \end{figure}

In addition to the phaseogram and long term light-curve, the SED for P1 and P2 is shown in Fig.~\ref{sed}. Both peaks are well described by a PWL model ($d\phi/dE = \phi_{0} (E/E_{0})^{-\Gamma} $) between 20 GeV and 700 GeV. The fit results are summarized in Table \ref{table:fitting_sed}. The reference energy $E_{0}$ was set to the decorrelation energy for each peak, defined as the energy that minimizes the correlation between the normalization flux $\phi_{0}$ and the rest of the parameters of the model (see Eq. (1) in \citeauthor{Abdo_2009_dec} \citeyear{Abdo_2009_dec}). The spectral index of P2 ($\Gamma_{2}$=3.03 $\pm$ 0.09) is considerably harder than the one for P1 ($\Gamma_{1}$=3.44 $\pm$ 0.15), while the flux for P1 is slightly larger below 30 GeV. These results are consistent with the most recent results from MAGIC \citep{MAGIC_teraelectronvolt, Ceribella_thesis} and VERITAS \citep{veritas_2011, Nguyen_2016}. Moreover, we confirm the PWL extension of P2 found by MAGIC and VERITAS above 500 GeV.

\begin{table}
    \caption{Fitted parameters of the spectral model with their statistical uncertainties, for each region. The reference energy $E_{0}$ was set to the decorrelation energy in each case.}
    \centering
    \begin{tabular}{cccc}
        \hline\hline
        \textbf{Region} & $E_{0}$ (GeV) & $\phi_{0}$ \ (cm$^{-2}$ s$^{-1}$ TeV$^{-1})$  & $\Gamma$  \\
        \hline
        P1              & 30                & (3.7 $\pm$ 0.4) $\cdot$ 10$^{-9}$  & 3.44 $\pm$ 0.15           \\
        \hline
        P2              & 40                & (1.56 $\pm$ 0.14) $\cdot$ 10$^{-9}$ & 3.03 $\pm$ 0.09  \\
        \hline
        P1 + P2              & 40                & (2.96 $\pm$ 0.20) $\cdot$ 10$^{-9}$ & 3.20 $\pm$ 0.08  \\
        \hline
        Bridge$_{M}$              & 30                & (8.1 $\pm$ 1.4) $\cdot$ 10$^{-9}$ & 3.5 $\pm$ 0.4  \\
        \hline
        Bridge$_{E}$              & 40   & (9.3 $\pm$ 2.5) $\cdot$ 10$^{-10}$ & 3.3 $\pm$ 0.6 \\
        \hline
        \end{tabular}
    \label{table:fitting_sed}
\end{table}

Pulsar analysis does not suffer from the systematic uncertainties in the background estimation that dominated the study of the Crab Nebula performance \citep{lst1_performance}. Thus, it is possible to quantify additional systematic uncertainties of the telescope with the pulsar signal. We tested different parameters in the analysis such as the cut efficiencies or the zenith angle and intensity cuts. We also shifted the true energy of the MC (up to 10\%), and computed the modified IRFs and spectra to test for a possible bias in the energy reconstruction. Additionally, we compared the SED of different sub-samples in the analysis. Adding all the contributions, the systematic uncertainties in the reconstruction of the spectral index for P1 and P2 is $\sim$0.34 and $\sim$0.21, while the uncertainties in the fluxes rise to $\sim$45\% and $\sim$20\%, respectively. These numbers are compatible with the ones found in the Crab Nebula study with the LST-1 above 60 GeV. 

To estimate the analysis energy threshold we used the MC simulations, weighing their spectrum by the one found in the Crab pulsar. These simulations were analyzed using the same analysis chain as for the observations. The peak of the true energy distribution of the MC events gives the energy threshold (E$_{\rm th}$) of the analysis, which depends on the Zd. For the LST-1 at Zd = 10 deg, the threshold estimated from the MC energy distribution is  E$_{\rm th}$=(18 $\pm$1) GeV, while at Zd = 32 deg it increases to E$_{\rm th}$=(29 $\pm$1) GeV. We estimate that for a spectrum similar to that of the Crab pulsar, the energy threshold below 35 degrees is $\approx$ 20 GeV.

%--------------------------------------------------------------------------
%
%
\subsection{Joint {\it Fermi}-LAT and LST-1 SED of the peaks}

\begin{table*}
\centering
\caption{Results of the best fits to the LST-1 and {\it Fermi}-LAT data for each peak and model. The fit was performed in the energy range from 100 MeV to 450 GeV (for P1) and 700 GeV (for P2). The statistical tests used to compare the models (i.e. AIC and BIC) are also shown. The reference energy was fixed to $E_{0}$=1 GeV for all the cases.}
\begin{tabular}{cccccccccc}

\hline
\textbf{SmoothBPWL}        &   $\phi_{0}$ (cm$^{-2}$ s$^{-1}$ TeV$^{-1}$)   & $\Gamma_{1}$       & $\Gamma_{2}$         & $E_{b}$ (GeV)  & $\gamma$ & $-2logL$ & AIC & BIC         \\ \hline
P1                                   & (1.27 $\pm$ 0.06) $\cdot$ 10$^{-4}$ & 1.811 $\pm$ 0.013 & 4.09 $\pm$ 0.20     & 6.8 $\pm$ 1.5   & 3.0 $\pm$ 0.4 & 25.5 & 35.5 & 42.8     \\ \hline
P2                              & (3.21 $\pm$ 0.20) $\cdot$ 10$^{-5}$  & 1.97 $\pm$ 0.03 & 3.15 $\pm$ 0.11    & 4.9 $\pm$ 0.9    & 1.1 $\pm$ 0.3  & 33.2 & 43.2 & 50.5        \\ \hline
\hline
\textbf{ExpCutPWL} &  $\phi_{0}$ (cm$^{-2}$ s$^{-1}$ TeV$^{-1}$)  & $\Gamma$        & $\lambda$ (GeV$^{-1}$) & $\beta$ (10$^{-1}$)    &   &   $-2logL$ & AIC & BIC   \\ \hline
P1                                   & (4.5 $\pm$ 0.4) $\cdot$ 10$^{-4}$    & 1.562 $\pm$ 0.015 & 6.0 $\pm$ 0.9     & 3.58 $\pm$ 0.08 &  &   35.9 & 43.9 & 49.8            \\ \hline
P2                                 & (2.8 $\pm$ 0.7)$\cdot$  10$^{-4}$  & 1.58 $\pm$ 0.03  & 29 $\pm$ 16     & 2.56 $\pm$ 0.12   & & 57.2 & 65.2 & 71.1               \\ \hline
Bridge$_{M}$                             & (2.1 $\pm$ 0.4) $\cdot$ 10$^{-4}$    & 1.42 $\pm$ 0.05 & 1.3 $\pm$ 0.5    & 4.5 $\pm$ 0.3 &    & 34.4 &     42.4      &    47.1  \\ \hline
Bridge$_{E}$                             & (2.1 $\pm$ 0.6) $\cdot$ 10$^{-5}$  & 1.16 $\pm$ 0.09  & 0.8 $\pm$ 0.5    & 5.1 $\pm$ 0.6   & &31.2 &   39.2  &      43.9   \\ \hline

\end{tabular}
\tablefoot{The fits for the bridge region were done between 200 MeV and 200 GeV (see text). Only the sub-exponential cutoff model was successfully fit. }
\label{table:joint_fit}
\end{table*}

Precise measurements at tens of GeV, in the energy range overlapping between {\it Fermi}-LAT and IACT, are needed to study the existence of spectral cuts or other spectral components. In this work, a joint fit with both LST-1 and {\it Fermi}-LAT data was performed between 100 MeV and 450 GeV for P1 and up to 700 GeV for P2. Two models were tested. The first model is a smooth broken PWL (SmoothBPWL, Eq. 1), and the second one is a typical PWL with a sub-exponential cutoff (ExpCutPWL, Eq. 2):
\begin{equation}
    \frac{d\phi}{dE} = \phi_{0}\left(\frac{E}{E_{0}}\right)^{-\Gamma_{1}}\left(1+\left(\frac{E}{E_b}\right)^{\frac{\Gamma_{2}-\Gamma_{1}}{\gamma}}\right)^{-\gamma} \ \ (\textrm{SmoothBPWL})
    \end{equation}
    \begin{equation}
    \frac{d\phi}{dE} = \phi_{0}\left(\frac{E}{E_{0}}\right)^{-\alpha} \exp\left(-(\lambda E)^{\beta}\right) \ \ (\textrm{ExpCutPWL})
\end{equation}

\begin{figure}[!htp]
    \centering
    \includegraphics[width=\hsize]{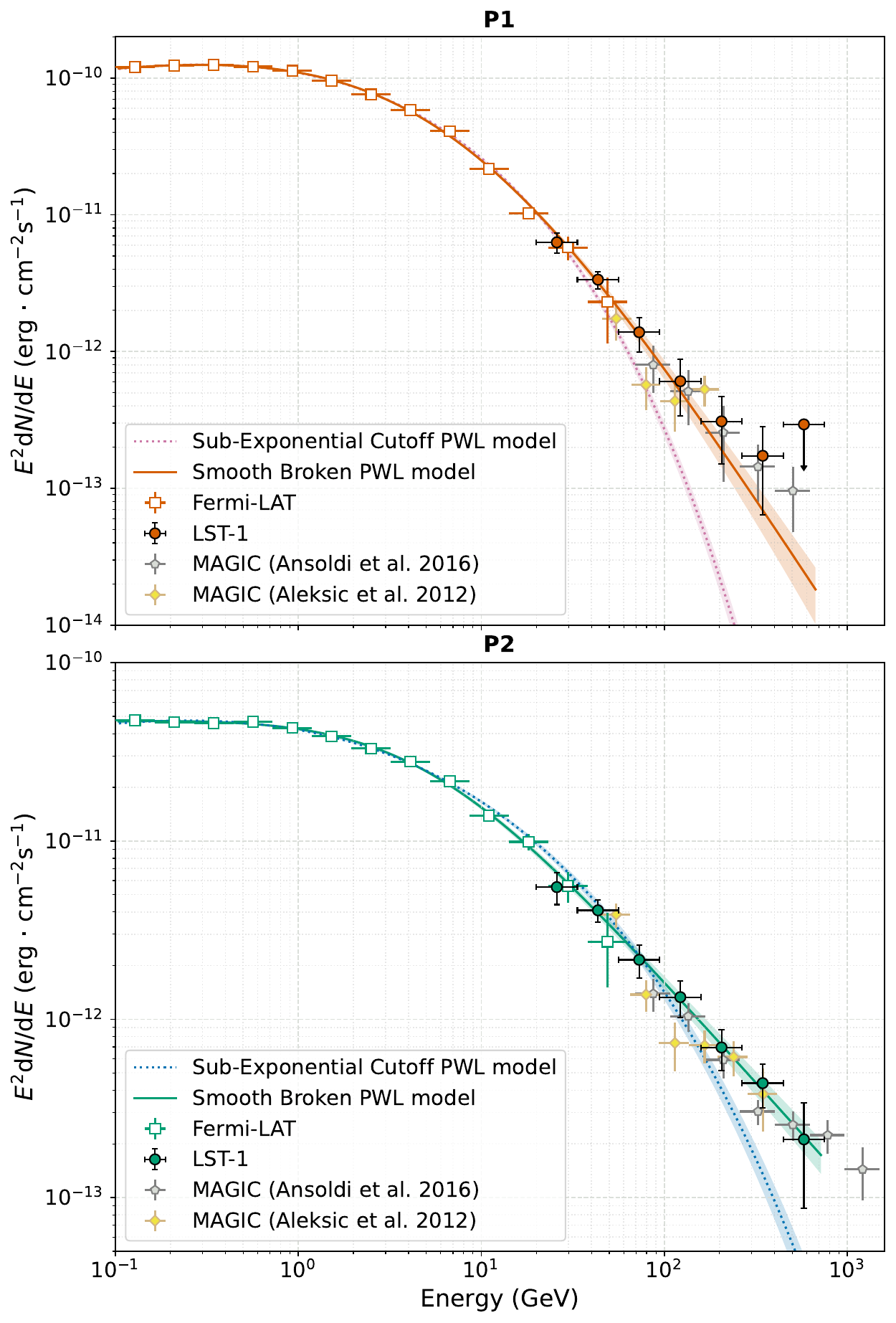}
    \caption{SED and joint fit using {\it Fermi}-LAT and LST-1 data from 100 MeV to 700 GeV for both P1 and P2 of the Crab pulsar. The points published by MAGIC working in stereo are shown as well.}
    \label{joint-fit}
\end{figure}

Both models were fit using a forward folding algorithm. The results of the fits together with the spectral points for P1 and P2 are shown in Fig.~\ref{joint-fit} and summarized in Table \ref{table:joint_fit}. MAGIC spectral points are also shown for comparison. A smooth transition between the instruments is clear and the spectral points of the LST-1 are compatible with the {\it Fermi}-LAT and the MAGIC points. The low statistical uncertainties of the LST-1 spectral points show that the telescope can fill the region between 20 GeV and 50 GeV with higher statistics than previous works from MAGIC ($\sim$60 hours analyzed in \citeauthor{Aleksi__2011} \citeyear{Aleksi__2011}).

The goodness of the fit for both models was compared using two statistics. The first one is the Akaike information criterion (AIC), defined as $\textrm{AIC}=2k-2\log{L}$ where $k$ is the number of free parameters and $L$ is the likelihood of the model. The second one is the Bayesian information criterion (BIC) defined as $\textrm{BIC}=k\ln{n}-2\log{L}$ where $n$ is the size of the sample. Both are information criteria that do not allow us to compute a p-value but to recognize in a qualitative way which model agrees better with the data. The smooth broken PWL shows, in general, a lower AIC and BIC value, with a difference of more than 5 in all the cases. For P1 this difference is $\Delta(\textrm{AIC})_{\rm P1}$=8.4 and $\Delta(\textrm{BIC})_{\rm P1}$=7.0; in the case of P2 the differences raise up to $\Delta(\textrm{AIC})_{\rm P2}$=22.0 and $\Delta(\textrm{BIC})_{\rm P2}$=20.6. This points to the smooth broken PWL as the preferred model to describe the observed fluxes. This is also observed in Fig.~\ref{joint-fit}, where although both fits seem to fit well in the low-energy spectrum, the very-high-energy spectral points agree better with the smooth broken PWL.

%--------------------------------------------------------------------------
%
%
\subsection{SED of the bridge emission}

\begin{figure}[!t]
    \centering
    \includegraphics[width=\hsize]{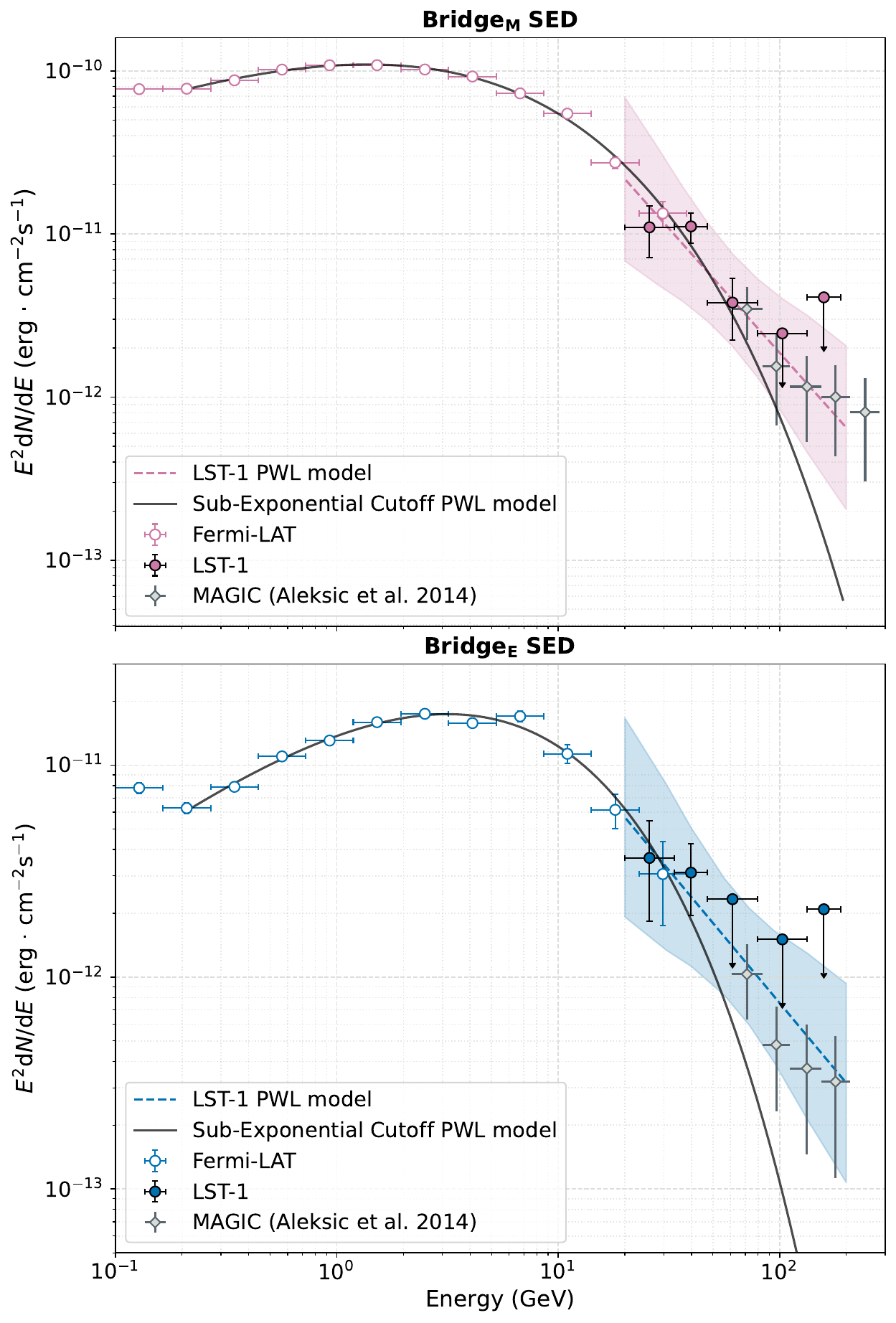}
    \caption{LST-1 SED from 20 GeV to 200 GeV for the two definitions of the bridge emission of the Crab pulsar. {\it Fermi}-LAT + LST-1 joint-fit to the Sub-Exponential Cutoff PWL model is shown in solid line. The points published by MAGIC working in stereo are shown as well. }
    \label{bridge}
\end{figure}

%We studied the bridge emission of the Crab pulsar and we computed the SED for each definition. 
The SED of the Crab pulsar bridge emission for each of the bridge regions defined in Sect. 4.1.1 is shown in Fig.~\ref{bridge}. The LST-1 SED  was fitted to a PWL between 20 GeV and 200 GeV. The results are shown in Table 3, the spectral indexes are $\Gamma_{M}$=(3.5 $\pm$ 0.4) and $\Gamma_{E}$=(3.3 $\pm$ 0.6). The LST-1 flux points are compatible with those reported by the MAGIC collaboration \citep{crabpulsar_magic_2014}, which extend up to 200 GeV. Above 100 GeV for the LST-1 the significance of the flux points is lower than for MAGIC and only upper limits can be calculated.

In addition, we did a joint fit using {\it Fermi}-LAT data and \LST{} from 200 MeV to 200 GeV. Data below 200 MeV were excluded from the fit because the analysis at the lowest energies led to unreliable flux estimation as indicated in Fig \ref{bridge}. In this case, since the signal above 100 GeV drops fast, we could only fit successfully the sub-exponential cutoff PWL model (shown in the solid line in Fig.~\ref{bridge}). The results are shown in Table 3. Although there is a hint of a PWL extension for both definitions, the lack of statistics prevents us from confirming or rejecting the existence of a cutoff in the bridge spectra. 

%%%%%%%%%%%%%%%%%%%%%%%%%%%%%%%%%%%%%%%%%%%%%%%%%%%%%%%%%%%%%%%%%%%%%%%%%%%%%%%%%
%
%
\section{Discussion and conclusions}

In this work, we have reported a detailed analysis of the first VHE gamma-ray pulsar detected with the LST-1. The results show that the LST-1 can detect the signal of the Crab pulsar at a high significance level and reconstruct its SED from 20 GeV up to 450 GeV for P1 and 700 GeV for P2.
%, making clear the great performance of one single LST to study pulsars. 
Both peaks P1 and P2 are significantly detected ($>$10$\sigma$) in this analysis. The VHE gamma-ray SED of each peak is well reproduced by a PWL compatible with previous results from the literature. P1 shows a softer spectrum ($\Gamma_{1}= 3.44 \pm 0.15$) than P2 ($\Gamma_{2}= 3.03 \pm 0.09$). Both peaks show similar fluxes ($\sim 3.5\cdot 10^{-9}$cm$^{-2}$ s$^{-1}$ TeV$^{-1}$) at $E=30$ GeV. The bridge emission is also significantly detected and the resulting spectra, for both definitions, are well described by PWLs with spectral indices $\Gamma_{\rm E}= 3.3\pm 0.6$ and $\Gamma_{\rm M}= 3.5\pm 0.4$ for Bridge$_{\rm E}$ and Bridge$_{\rm M}$ respectively. The long-term light curve of the Crab pulsar using the LST-1 data was also studied over three different years, finding a value of $\chi^2=12.8/$10  when the pulsed integral flux is fit to a constant, demonstrating the stability of the overall energy released in the pulsed signal over time within our statistical uncertainties. 

%The consistency of the LST-1 results in the overlapping region with {\it Fermi}-LAT proves its performance at E$<$50 GeV. Since we can study this region with high statistics, we performed robust joint fits to study the overall emission of gamma rays.  
%For both P1 and P2 pulsed emission, it is clear that the transition is very smooth between the instruments, rejecting for both peaks the existence of a cut-off at a few GeV.
%For both P1 and P2 pulsed emission, it is clear that the transition is very smooth between the instruments, rejecting for both peaks the existence of a cut-off at a few GeV.
The consistency of the LST-1 results in the overlapping region with {\it Fermi}-LAT proves its performance at E$<$50 GeV. 
We perform joint {\it Fermi}-LAT and LST-1 fits to study the overall gamma-ray emission of the Crab pulsar over four decades of energy.
For both P1 and P2, there is a smooth transition in the SED measured by the two instruments, rejecting the existence of additional spectral components in the overlapping energy range. 
We find that both spectra are well described by smooth broken power laws.
For the bridge emission, the lack of statistics above 100 GeV makes it difficult to statistically reject the sub-exponential cut-off. 

The results presented in this paper confirm the Crab pulsar as a TeV lepton accelerator. As described in \citet{MAGIC_teraelectronvolt}, the radiation produced by the most energetic electrons and positrons cannot originate via synchro-curvature processes even if the lower energy ones could have a different origin. Although there are current models that predict that the emission could be generated by multiple particle populations \citep{Harding_2021}, we see a smooth transition between {\it Fermi}-LAT and LST-1 data that could point towards the measured emission being produced by a single population of electrons. Thus, a plausible origin of the gamma-ray emission remains the inverse Compton scattering of ambient photon fields or the synchrotron-self Compton from pairs. The exact acceleration place for the electrons and positrons remains unclear and locations like the outer gap \citep{2013_Hirotani}, slot gap \citep{2007_Harding}, the pulsar striped wind \citep{2012_Petri} or narrow zones outside the light cylinder \citep{Aharonian_2012} continue to be plausible regions. Variability studies were performed in X-rays \citep{2016_Ge}, indicating a time evolution in the pulse profile at those energies. To the extent detectable by current sensitivity, the long-term light curve of the Crab pulsar does not show signs of variability in the LST-1 data, the same happening with its accompanying nebula \citep{lst1_performance}. 

Being able to study the Crab pulsar in such detail with the first LST telescope indicates a significant improvement in sensitivity over the previous generation of IACTs. 
%This is very encouraging since it points to the possible detection of more pulsars above 20 GeV in the near future (although some works predict low fluxes above 50 GeV; \citeauthor{2015_McCann} \citeyear{2015_McCann}), especially when the next three LSTs, currently under construction in La Palma, become available to work as a stereoscopic system, hence improving the sensitivity below 100 GeV by approximately one order of magnitude. 
This points to the possible detection of more pulsars above 20 GeV in the near future (although some work predicts low fluxes above 50 GeV; \citeauthor{2015_McCann} \citeyear{2015_McCann}), especially when the next three LSTs, currently under construction in La Palma, become available. The four LSTs operating as a stereoscopic system will improve the current sensitivity below 100 GeV by about an order of magnitude.
The discovery of VHE emission from other pulsars would open new possibilities for the study of gamma-ray emission from these objects.

%%%%%%%%%%%%%%%%%%%%%%%%%%%%%%%%%%%%%%%%%%%%%%%%%%%%%%%%%%%%%%%%%%%%%%%%%%%%%%%%%
%
%
\begin{acknowledgements}
We gratefully acknowledge financial support from the following agencies and organisations:\\ \\

Conselho Nacional de Desenvolvimento Cient\'{\i}fico e Tecnol\'{o}gico (CNPq), Funda\c{c}\~{a}o de Amparo \`{a} Pesquisa do Estado do Rio de Janeiro (FAPERJ), Funda\c{c}\~{a}o de Amparo \`{a} Pesquisa do Estado de S\~{a}o Paulo (FAPESP), Funda\c{c}\~{a}o de Apoio \`{a} Ci\^encia, Tecnologia e Inova\c{c}\~{a}o do Paran\'a - Funda\c{c}\~{a}o Arauc\'aria, Ministry of Science, Technology, Innovations and Communications (MCTIC), Brasil;
Ministry of Education and Science, National RI Roadmap Project DO1-153/28.08.2018, Bulgaria;
Croatian Science Foundation, Rudjer Boskovic Institute, University of Osijek, University of Rijeka, University of Split, Faculty of Electrical Engineering, Mechanical Engineering and Naval Architecture, University of Zagreb, Faculty of Electrical Engineering and Computing, Croatia;
Ministry of Education, Youth and Sports, MEYS  LM2015046, LM2018105, LTT17006, EU/MEYS CZ.02.1.01/0.0/0.0/16\_013/0001403, CZ.02.1.01/0.0/0.0/18\_046/0016007 and CZ.02.1.01/0.0/0.0/16\_019/0000754, Czech Republic; 
CNRS-IN2P3, the French Programme d’investissements d’avenir and the Enigmass Labex, 
This work has been done thanks to the facilities offered by the Univ. Savoie Mont Blanc - CNRS/IN2P3 MUST computing center, France;
Max Planck Society, German Bundesministerium f{\"u}r Bildung und Forschung (Verbundforschung / ErUM), Deutsche Forschungsgemeinschaft (SFBs 876 and 1491), Germany;
Istituto Nazionale di Astrofisica (INAF), Istituto Nazionale di Fisica Nucleare (INFN), Italian Ministry for University and Research (MUR);
ICRR, University of Tokyo, JSPS, MEXT, Japan;
JST SPRING - JPMJSP2108;
Narodowe Centrum Nauki, grant number 2019/34/E/ST9/00224, Poland;
The Spanish groups acknowledge the Spanish Ministry of Science and Innovation and the Spanish Research State Agency (AEI) through the government budget lines PGE2021/28.06.000X.411.01, PGE2022/28.06.000X.411.01 and PGE2022/28.06.000X.711.04, and grants PID2022-139117NB-C44, PID2019-104114RB-C31,  PID2019-107847RB-C44, PID2019-104114RB-C32, PID2019-105510GB-C31, PID2019-104114RB-C33, PID2019-107847RB-C41, PID2019-107847RB-C43, PID2019-107847RB-C42, PID2019-107988GB-C22, PID2021-124581OB-I00, PID2021-125331NB-I00, PID2022-136828NB-C41, PID2022-137810NB-C22, PID2022-138172NB-C41, PID2022-138172NB-C42, PID2022-138172NB-C43, PID2022-139117NB-C41, PID2022-139117NB-C42, PID2022-139117NB-C43, PID2022-139117NB-C44, PID2022-136828NB-C42 funded by the Spanish MCIN/AEI/ 10.13039/501100011033 and “ERDF A way of making Europe; the ``Centro de Excelencia Severo Ochoa" program through grants no. CEX2019-000920-S, CEX2020-001007-S, CEX2021-001131-S; the ``Unidad de Excelencia Mar\'ia de Maeztu" program through grants no. CEX2019-000918-M, CEX2020-001058-M; the ``Ram\'on y Cajal" program through grants RYC2021-032991-I  funded by MICIN/AEI/10.13039/501100011033 and the European Union “NextGenerationEU”/PRTR; RYC2021-032552-I and RYC2020-028639-I; the ``Juan de la Cierva-Incorporaci\'on" program through grant no. IJC2019-040315-I and ``Juan de la Cierva-formaci\'on"' through grant JDC2022-049705-I. They also acknowledge the ``Atracción de Talento" program of Comunidad de Madrid through grant no. 2019-T2/TIC-12900; the project ``Tecnologi\'as avanzadas para la exploracio\'n del universo y sus componentes" (PR47/21 TAU), funded by Comunidad de Madrid, by the Recovery, Transformation and Resilience Plan from the Spanish State, and by NextGenerationEU from the European Union through the Recovery and Resilience Facility; the La Caixa Banking Foundation, grant no. LCF/BQ/PI21/11830030; Junta de Andaluc\'ia under Plan Complementario de I+D+I (Ref. AST22\_0001) and Plan Andaluz de Investigaci\'on, Desarrollo e Innovaci\'on as research group FQM-322; ``Programa Operativo de Crecimiento Inteligente" FEDER 2014-2020 (Ref.~ESFRI-2017-IAC-12), Ministerio de Ciencia e Innovaci\'on, 15\% co-financed by Consejer\'ia de Econom\'ia, Industria, Comercio y Conocimiento del Gobierno de Canarias; the ``CERCA" program and the grants 2021SGR00426 and 2021SGR00679, all funded by the Generalitat de Catalunya; and the European Union's ``Horizon 2020" GA:824064 and NextGenerationEU (PRTR-C17.I1). This research used the computing and storage resources provided by the Port d’Informació Científica (PIC) data center.
State Secretariat for Education, Research and Innovation (SERI) and Swiss National Science Foundation (SNSF), Switzerland;
The research leading to these results has received funding from the European Union's Seventh Framework Programme (FP7/2007-2013) under grant agreements No~262053 and No~317446;
This project is receiving funding from the European Union's Horizon 2020 research and innovation programs under agreement No~676134;
ESCAPE - The European Science Cluster of Astronomy \& Particle Physics ESFRI Research Infrastructures has received funding from the European Union’s Horizon 2020 research and innovation programme under Grant Agreement no. 824064.

\end{acknowledgements}

\section*{Author contribution}
A. Mas-Aguilar: project coordination, LST-1 Crab pulsar analysis (phaseogram, long term light-curve and SED), joint fit LST-1 and Fermi-LAT, systematic error estimation, results discussion.  R. López-Coto: project coordination, LST-1 data analysis, Crab pulsar analysis (phaseogram), cross-check on spectral analysis, results discussion. M. López-Moya: project coordination, Fermi-LAT analysis of the Crab pulsar, results discussion. L. Gavira: LST-1 spectral analysis cross-check, study of systematic errors. All corresponding authors have participated in the paper drafting and edition. The rest of the authors have contributed in one or several of the following ways: design,construction, maintenance and operation of the instrument(s) used to acquire the data; preparation and/or evaluation of the observation proposals; data acquisition, processing, calibration and/or reduction; production of analysis tools and/or related Monte Carlo simulations; discussion and approval of the contents of the draft.


% for the bibliography, at the end
\bibliographystyle{aa} % style aa.bst
\bibliography{Bibliography} % your references Yourfile.bib

\end{document}